\documentclass[twocolumn,twoside,slac_two]{revtex4}
\usepackage{graphicx}
\usepackage{fancyhdr}
\usepackage{color}
\usepackage{amsmath}
\usepackage{hql}
\usepackage[printonlyused,nohyperlinks]{acronym}
\usepackage{ifpdf} 
           \ifpdf
             \makeatletter
               \g@addto@macro\Gin@extensions{,.eps}
             \makeatother
           \fi
\graphicspath{{.}{./figures/}}
\begin{document}

\title{The \acs*{T2K} Neutrino Oscillation Experiment
and Possible Future Projects}

\author{N. C. Hastings}
\affiliation{The University of Tokyo}

\begin{abstract}
  The \ac*{T2K} experiment is a next generation long baseline neutrino
  oscillation experiment utilising the \ac*{JPARC} high intensity
  proton synchrotron.  After a brief introduction of the current
  understanding of neutrino mixing, the \ac*{T2K} experiment, its
  current status and the expected physics results are presented.
  Then, possibilities for future neutrino oscillation experiments
  utilising \ac*{JPARC} are discussed.
\end{abstract}
\acresetall
\maketitle

\thispagestyle{fancy}

\section{Introduction}
Neutrino oscillations have been observed via their disappearance signature in
atmospheric and long baseline accelerator produced \num and solar and reactor
produced \nue and \nueb. These oscillations can be readily explained
by the \ac*{MNS} mixing matrix which relates the neutrino mass Eigen
states to the flavour Eigen states:
\begin{equation}
  \begin{split}
    \overset{\mbox{Flavour}}{
      \begin{pmatrix} \nue \\ \num \\\nut \end{pmatrix}
    }
    =&
    \underbrace{
      \begin{pmatrix}
        1 & 0 & 0\\
        0 & c_{23} & s_{23}\\
        0 & -s_{23}& c_{23}
      \end{pmatrix}
    }_{\mbox{Atmo's \& Long BL}}
    \overbrace{
      \begin{pmatrix}
        c_{13} & 0 & s_{13}e^{-i\delta}\\
        0 & 1 & 0\\
        -s_{13}e^{i\delta}& 0 & c_{13}
      \end{pmatrix}
    }^{\mbox{Unknown}}\\
    &\times
    \underbrace{
      \begin{pmatrix}
        c_{12} & s_{12} & 0\\
        -s_{12} & c_{12} & 0\\
        0 & 0 & 1
      \end{pmatrix}
    }_{\mbox{KamLAND \& Solar}}
    \overset{\mbox{Mass}}{
      \begin{pmatrix} \nu_1 \\ \nu_2 \\\nu_3 \end{pmatrix}
    },\label{eq:mns}
  \end{split}
\end{equation}
where $s_{ij}\equiv\sin\theta_{ij}$ and $c_{ij}\equiv\cos\theta_{ij}$.
The KamLAND experiment observing \nueb from reactors and observations
of deficits of solar neutrinos provide measurements of \sinstt{12} and
the mass difference \dms{12}.  Observations of atmospheric \num
deficits and angular distributions and long baseline accelerator
produced neutrino experiments have provided measurements of
\sinstt{23} and the mass difference squared \dms{23}. There are still
significant unanswered questions.
The world average for \sinstt{23} is consistent with unity: but is
this it's true value? I.e. is the mixing maximal? The \ac*{T2K} experiment
will address this question by refining the \num disappearance analysis
of earlier experiments providing a precision measurement of \sinstt{23}.
The CHOOZ reactor experiment tells us that \sinstt{13} is small
($<0.13$, 90\% CL), but is it really non-zero? \ac*{T2K} will endeavour to discover a
non-zero \sinstt{13} by searching for \decay{\num}{\nue} oscillations.

\section{\acs*{T2K}}
\subsection{Neutrino Beam}
The \ac*{T2K} experiment is a ``Next generation'' long baseline neutrino
oscillation experiment. A \num beam will be produced using the newly
constructed proton synchrotron at \ac*{JPARC} in Tokai on the East
coast of Japan. The neutrinos will propagate 295~\kilo\metre\ through
the earth to Kamioka where they will be detected by the existing
\ac*{SK} water Cerenkov detector. The experiment is scheduled
to start in April 2009 and is planned to run for five years with a beam
power of 0.75~\mega\watt. This represents approximately a factor of
100 increase in data sample of the \ac*{K2K} experiment.

The neutrino beam will be created by striking a carbon target with an
intense proton beam producing pions. A series of three magnetic horns
selects and focuses the positive pions into a 110~\metre\ helium
filled decay volume --- where they decay --- producing \muplus and
\num. Just downstream of the decay volume the muon beam's intensity,
profile and direction are monitored providing information on the same
parameters for the accompanying neutrinos.

\begin{figure*}[htbp]
  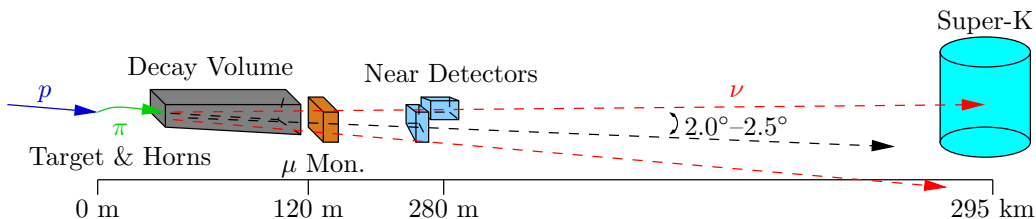
  \caption{\protect\label{fig:t2kcartoon}Configuration of the \ac*{T2K} neutrino beam indicating the beam monitors and detectors and the off-axis configuration.}
\end{figure*}

\begin{figure}[htbp]
  \resizebox{0.5\columnwidth}{!}{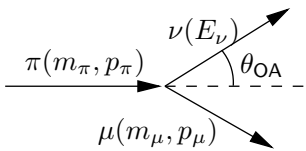}
  \caption{\protect\label{fig:pidecay}Pion decay kinematics in the laboratory frame.}
\end{figure}

\begin{figure}[htbp]
  \includegraphics[height=!,width=0.9\columnwidth]{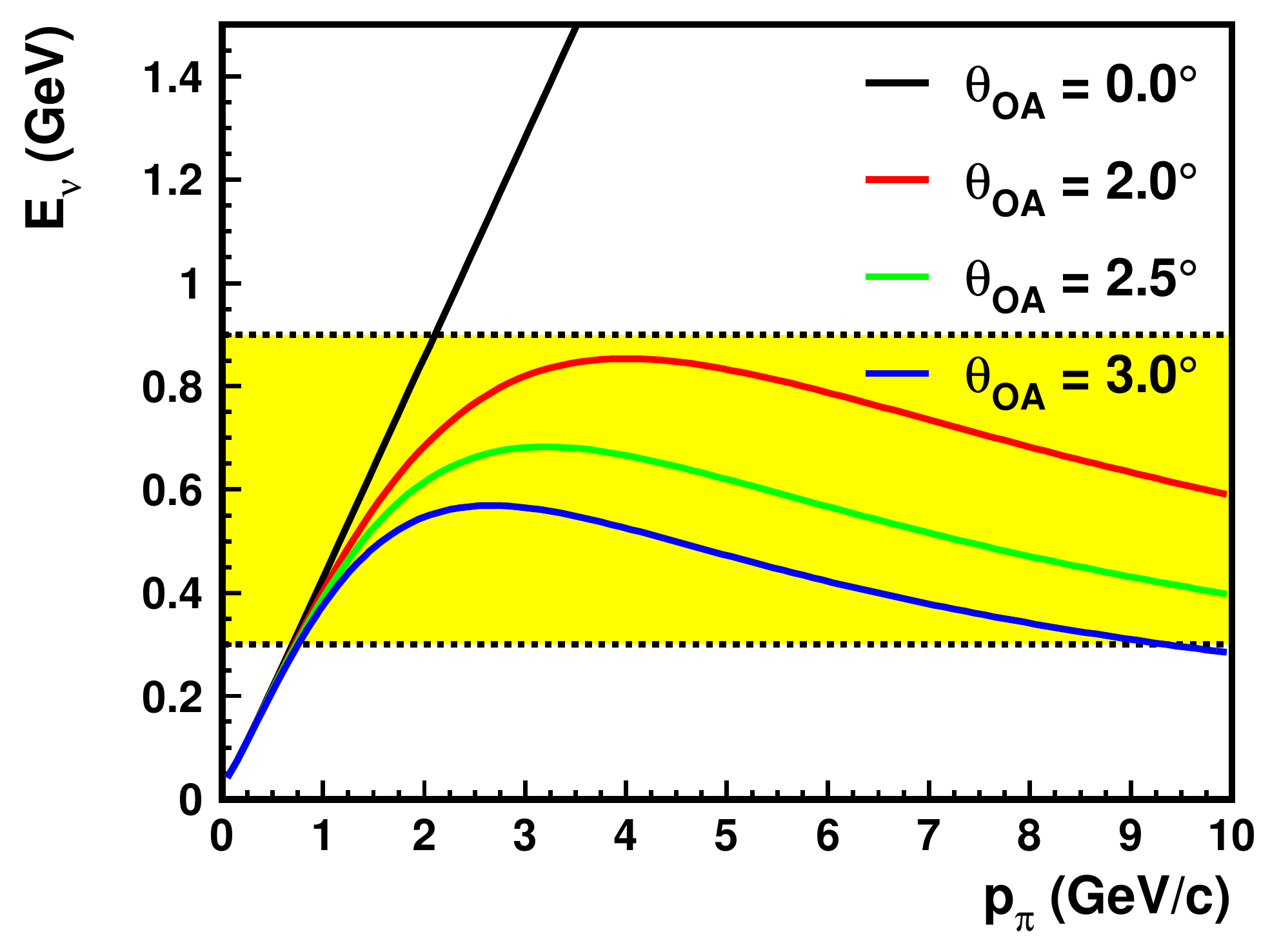}
  \caption{\label{fig:enu}Neutrino beam energy as a function of pion
    energy for different off axis angles.}
\end{figure}

Located two hundred and eighty metres from the target are a pair of
``Near Detectors''. These detectors are designed to monitor the
neutrino beam itself.  The neutrino beam then travels 295~\kilo\metre\
to the Super-Kaimokande detector.  A novel feature of the beam line is
that it utilises an ``off-axis'' configuration: the neutrino beam is
not aimed directly at Super-K, but 2.5\degree\ from it. A cartoon
of the configuration is shown in Fig.~\ref{fig:t2kcartoon}.
This configuration utilises a nice feature of the pion decay kinematics. The
neutrino energy is given by
\begin{equation}
  E_\nu = \frac{m_\pi^2-m_\mu^2}{2(E_\pi-p_\pi\cos\toa)},
  \label{eq:pidecay}
\end{equation}
where all parameters are in the laboratory frame and \toa
is the angle between the neutrino and pion flight directions. The
decay is illustrated in Fig.~\ref{fig:pidecay}.
Figure~\ref{fig:enu}
shows Eq.~\ref{eq:pidecay} plotted for different \toa
demonstrating how selecting neutrinos at non-zero angles from the pion
beam direction provides an almost monochromatic neutrino beam
energy. \ac*{T2K} is setup to run with \toa=2.5\degree\ which selects
600~\mev to 700~\mev neutrinos placing the first neutrino oscillation
maximum at Super-K for $\Delta m^2=2.5\times10^{-3}~\electronvolt\squared\!/c^4$.

\subsection{Schedule}
Commissioning of the \ac*{JPARC} linac, \ac*{RCS} and \ac*{MR} began
in 2006, 2007 and 2008 respectively. All components excluding the fast
extraction kickers of the \ac*{MR} were installed as of April 2008.
Closed orbit in the \ac*{MR} was established in May 2008 and the fast
extraction kicker installation is scheduled for June 2008. Acceleration
of the beam in the \ac*{MR} to 30~\gev and fast extraction to the
neutrino beamline are scheduled for December 2008 and April 2009.

\subsection{Apparatus}
The apparatus of the \ac*{T2K} experiment consist of four major parts:
The primary beamline, the secondary beamline, the near detectors and
the far detector.
\subsubsection{Primary Beamline}
This first section of the primary beamline focusses the protons from
the \ac*{MR} in preparation for the transportation through the
curved \ac*{SC} or arc section which bends the beam almost
90\degree\ towards the target. The \ac*{SC} magnets in the arc section
are combined dipole/quadrapole magnets consisting of 14
doublets. The final section of the primary beamline defocuses the beam
and delivers it to the target. The beamline incorporates 21 position,
19 profile, 5 intensity and 50 beam loss monitors.

\subsubsection{Secondary Beamline}
The secondary beamline consists of a helium vessel containing a graphite
target located in the first of three ``horn'' magnets designed to
focus the pions produced when the proton beam strikes the target. Horn
operation has been tested up to 320~\kilo\ampere. The facility is
complete with a fully remote crane and maintenance area to allow horn
and target replacement in the event of failure of any component. Civil
construction was finished in May 2008. Downstream of the target
station is a 110~\metre\ long helium filled decay volume which ends
with a hadron absorber consisting of water cooled graphite blocks,
behind which is an array of ionisation chambers and silicon PIN diodes
which monitor the direction and profile of the muons created from the
pion decays.
\subsubsection{Neutrino Detectors}
Two near detectors: on-axis and off-axis will be located in a pit
280~\metre\ downstream of the target. The on-axis detector, INGRID,
consists of a 10~\metre\ by 10~\metre\ configuration of stacks of
scintillators interleaved with iron sheets. This detector, to be
completed by April 2009, is designed to directly monitor the \num beam
profile and direction with a resolution of 0.18~\milli\radian.  The
off-axis detector, ND280, is designed to measure the neutrino flux in
the direction of \ac*{SK} for both \num and \nue utilising a fine
grained tracker for accurate reconstruction of \ac*{CCQE} events. The
detector will also be capable of measuring the relative \ac*{CCQE} to
\ac*{CCNQE} cross sections and will have calorimeters which will be
utilised to measure the neutral current $\pi^0$ production rate
(background source for the \decay{\num}{\nue} search). The \ac*{ND}
will be completed in the second half of 2009.

Located 295~\kilo\meter\ from \ac*{JPARC} is the far detector: the
22.5~\kilo\ton\ fiducial volume water Cerenkov detector
\ac*{SK}. \ac*{SK} was fully re-populated with \acp*{PMT} in 2006, now
with a total of 11146 and 1885 \acp*{PMT} in the inner and outer
volumes respectively and will have new electronics installed in
September 2008.

\subsection{Expected Physics Results}

The survival probability for the neutrinos produced in \ac*{T2K} is
given by 
\begin{equation}
  \pr{\num}{\num}\simeq 1-\sin^22\tmix{23}\sin^2
  \left(\frac{1.27\dms{23} L}{E_\nu}\right),\label{eq:pmm}
\end{equation}
where $L$ and $E_\nu$ are the neutrino flight length and energy.
As such the observed number of \num events at \ac*{SK} as a function
of $E_\nu$ for the fixed $L=295~\kilo\metre$ can be fitted to
Eq.(\ref{eq:pmm}) to extract \tmix{23} and \dms{23}.

\begin{figure}[htbp]
  \resizebox{0.6\columnwidth}{!}{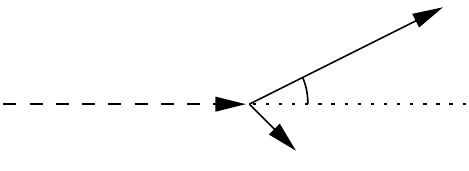}
  \caption{\protect\label{fig:ccqe}\acf*{CCQE} interaction}
\end{figure}

At \ac*{SK} the neutrino energy can be reconstructed though \ac*{CCQE}
interactions as illustrated in Fig. \ref{fig:ccqe}.
From this interaction in \ac*{SK}, $(E_\mu,\theta)$ are extracted and
used to calculate the incident \num energy from the kinematics of the
interaction:
\begin{equation}
  E_\nu =  \frac{m_n E_\mu - m_\mu^2/2}{m_n - E_\mu + p_\mu\cos\theta}.
\end{equation}

The expected number of neutrinos at the \ac*{ND} and \ac*{SK} can be
given by $ N_\mathrm{ND} = \phi_\mathrm{ND} \sigma_\mathrm{ND}$ and
$N_\mathrm{SK} = \phi_\mathrm{SK} \sigma_\mathrm{SK} P_\mathrm{osc}$
respectively. Of these $\sigma_\mathrm{SK}$, $\sigma_\mathrm{ND}$ and
$\phi_\mathrm{ND}$ are studied with the \ac*{ND}. Treating the fluxes
as a ratio: $R_{N/F} = \frac{\phi_\mathrm{SK}}{\phi_\mathrm{ND}}$, the
number of events can be predicted: $N_\mathrm{SK}^\mathrm{pred} =
N_\mathrm{ND}^\mathrm{obs}R_{N/F}\frac{\sigma_\mathrm{SK}}{\sigma_\mathrm{ND}}$. To
calculate the so called near-far ratio, $R_{N/F}$, the hadron
production in the \ac*{T2K} target needs to be well understood. This
will be studied with the \shine experiment (NA49 upgraded to NA61) at
\cern. \shine runs with a proton beam on carbon target with
conditions tuned to match those of \ac*{T2K}. The detector has good
tracking and particle identification enabling discrimination between
$\pi^\pm$ and $K^\pm$ and reconstruction of \kz.

The approved total integrated beam power to \ac*{T2K} is
$0.75\times1500~\hour$. It is expected that this will be delivered
within five years of operation. The following sensitivities assume
this integrated beam power.

\subsubsection{\num disappearance: Precision \tmix{23} and \dms{23}
  measurements} 
\begin{figure}[htbp]
  \includegraphics[height=!,width=0.9\columnwidth,trim= 0 50 340 50,clip]
      {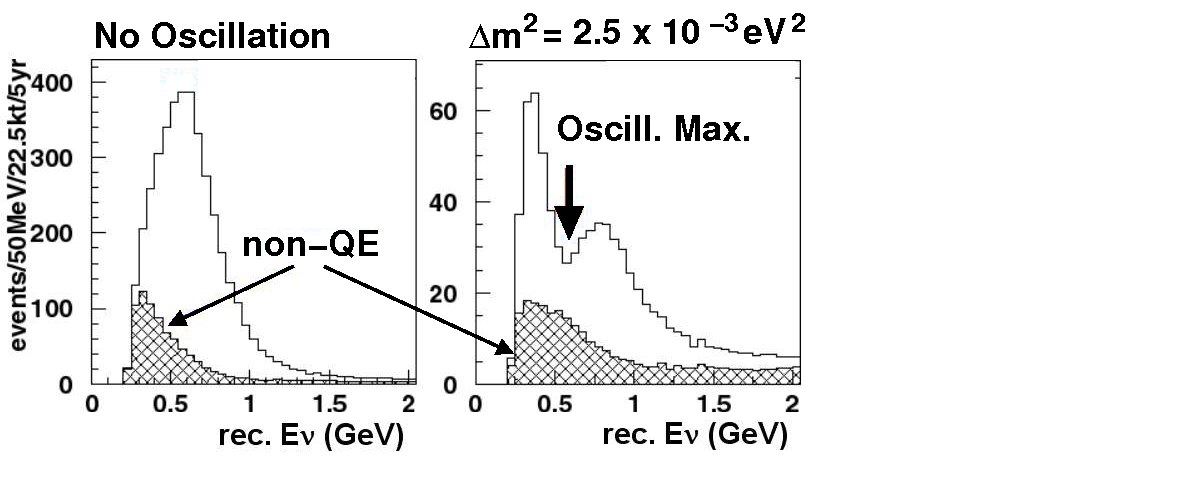}
      \caption{\label{fig:numudis}Reconstructed energy spectrums for
        \num disappearance analysis. The hatched areas show the
        NonQuasi-Elastic backgrounds. Left plot shows the case of no
        oscillations, the right plot shows oscillations with
        $\dms{23}=2.5\times10^{-3}~\electronvolt\squared$. The dip
        caused by the first oscillation maximum is indicated. The
        position of the dip is governed by \dms{23} and the
        depth by \sinstt{23}.}
\end{figure}

Figure~\ref{fig:numudis} shows the expected neutrino
energy spectrums and NonQuasi-Elastic backgrounds for the case of no
oscillations and oscillations with $\dms{23}=2.5\times 10^{-3}~\electronvolt\squared$.
Studies have shown that the systematic uncertainties in the flux
prediction can be controlled to better than 25\% resulting in the
final uncertainties in \tmix{23} and \dms{23} being statistically
limited. The resulting expected achieved sensitivities are
$\delta(\sin^22\tmix{23}) \simeq 0.01$ and $\delta(\dms{23}) <
10^{-4}~\electron\volt\squared$.
\subsubsection{\nue appearance: non-zero \tmix{13}}
Utilising the electron identification abilities of \ac*{SK}
with a background estimation error of 10\% \ac*{T2K} will be able
to place limits of $\sin^2 2\tmix{13} < 0.008$ (90\% C.L.) for:
$\dcp=0$, $\dms{13}=2.5\times10^{-3}~\electron\volt\squared$ and
$\sin^22\tmix{23}=1$. Or better than $\sin^2 2\tmix{13} < 0.02$ (90\%
C.L.) for any value of \dcp, representing a factor of ten improvement
of the CHOOZ limits.  Sensitivity contours for various values of
\sinstt{23} in the \dcp-\sinstt{13} plane are shown in
Fig.~\ref{fig:numunue}.
\begin{figure}[htb]
  \includegraphics[width=0.7\columnwidth,height=!]{flux04a_40gev_2\DOT
    5deg-s13-delta-q23_t12sub.pdf}
  \caption{\label{fig:numunue}Sensitivity contours in the
    \sinstt{13}-\dcp for various values of \sinstt{23}.}
\end{figure}

\section{Possible Future Projects}
If a significant \decay{\num}{\nue} signal is seen with \ac*{T2K}
there will be strong motivation carry out further studies utilising
the neutrino beamline at \ac*{JPARC}. The goals of such studies
include refining the \decay{\num}{\nue} appearance measurement
providing more precise information on \sinstt{13} and to try \CP
violation physics studies. To achieve such goals it will be necessary
increase the neutrino beam intensity and hence the proton beam
intensity. Additionally it will be vital to improve the far detector
in terms of target mass and/or technology and the baseline and beam
angle configuration.

\begin{figure}[htbp]
  \plotroot[0.8]{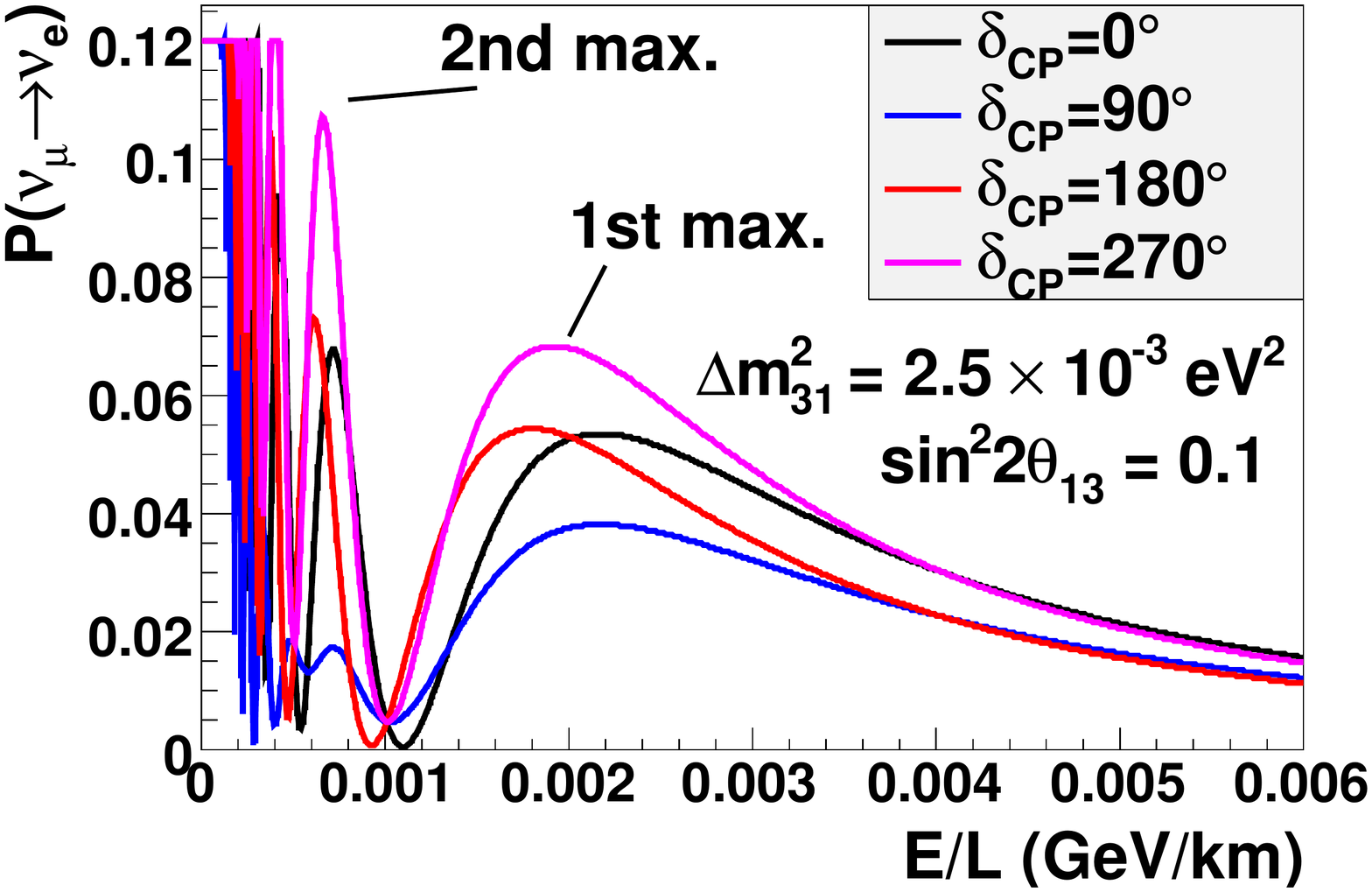}
  \caption{\label{fig:pmue}\pr{\num}{\nue} as a function of
    $E/L$ for different values of \dcp. The gross amplitude is
    determined by \sinstt{13}, but the value of \dcp changes the
    relative amplitude of the first and second maxima.}
\end{figure}

\begin{figure*}[bt]
  \vspace*{1\baselineskip}
  \includegraphics[width=1.0\textwidth,height=!]{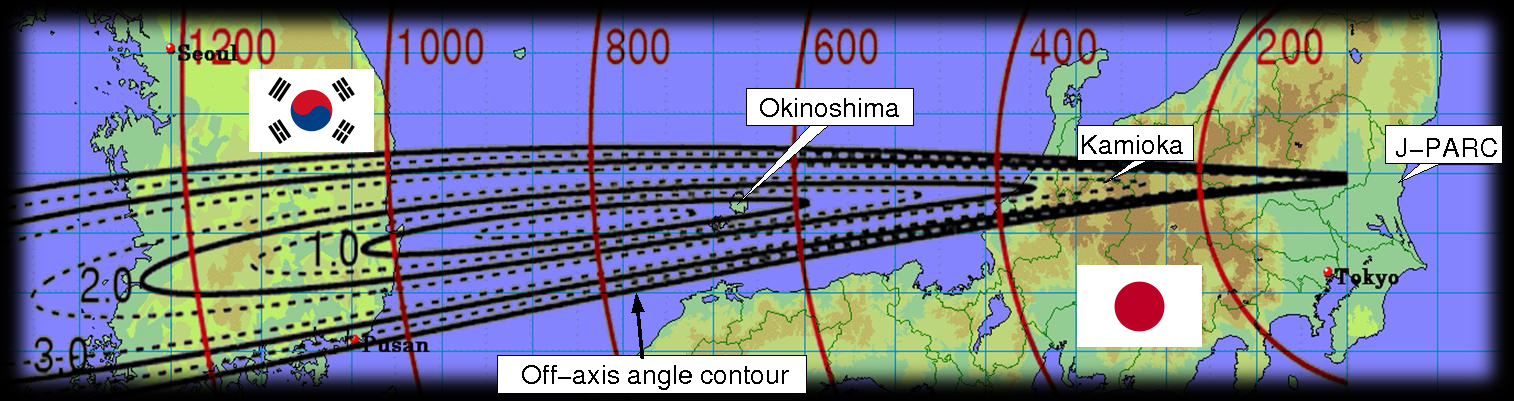}
  \caption{\label{fig:map}Map with off-axis angle and distance from
    production point contours for the \ac*{JPARC} neutrino beam.}
\end{figure*}

With the higher statistics from such an upgraded experiment it is
important to consider the \decay{\num}{\nue} transition probability in
some detail. An expansion of the analytic expression for
\pr{\num}{\nue} around the small parameters
$\alpha \equiv \frac{\dms{21}}{\dms{31}}$ and \sintt{13},
yields\cite{Akhmedov:2004ny}:
\begin{equation}
  \pr{\num}{\nue} = \sinstt{13}T_1
  + \alpha\sintt{13}\overbrace{(T_2-T_3)}^{\mbox{\smaller Interference}}
  +\alpha^2T_4.\label{eq:pnumunumu}
\end{equation}
Where:
\begin{equation}
  \begin{split}
    T_1 =& \sinst{23}\frac{\sin^2[(A-1)\Delta]}{(A-1)^2},
    \quad \leftarrow\mbox{Atmospheric} \\
     T_2=&
     \cos\dcp
     \sintt{12}\sintt{23}
     \cos\Delta\\
     &\hspace*{12ex}\times\frac{\sin(A\Delta)}{A}\frac{\sin[(A-1)\Delta]}{A-1},
     \\
     T_3=&
     \sin\dcp
     \sintt{12}\sintt{23}
     \sin\Delta\\
     &\hspace*{12ex}\times\frac{\sin(A\Delta)}{A}\frac{\sin[(A-1)\Delta]}{A-1},
     \\
     T_2-T_3=&
     \sintt{12}\sintt{23}
     \cos(\Delta+\dcp)\\
     &\hspace*{12ex}\times\frac{\sin(A\Delta)}{A}\frac{\sin[(A-1)\Delta]}{A-1},
     \\
     T_4=&\cosst{23}\sinstt{12}\frac{\sin^2(A\Delta)}{A^2}.
     \quad \leftarrow \mbox{Solar}
  \end{split}\label{eq:pnumunumu-terms}
\end{equation}
And:
\begin{equation}
  A \equiv \frac{2EV}{\dms{31}},\quad
  \Delta \equiv \frac{\dms{31}L}{4E},
\end{equation}
where $V$ is the potential seen by a neutrino passing through the earth
and as such $A$ is known as the ``matter effect'' term.  In the
representation of Eq.~\ref{eq:pnumunumu} and
Eq.~\ref{eq:pnumunumu-terms} it is instructive to consider the
contributing terms in the energy region of the \ac*{T2K}
neutrino beam, $\lesssim 1~\gev$. $T_1$, the ``atmospheric'' term is the
dominant component driving the oscillation with period about double
the length of the \ac*{T2K} baseline.  $T_4$, the ``solar'' term is
not related to \sintt{13} and has orders of magnitude longer period
and such plays no relevant part in the following. The $T_2$ and $T_3$
terms are 90\degree out of phase, and when combined to $T_2-T_3$, \dcp
introduces a phase offset to the oscillation which when combined into
the full oscillation probability can appear as a small change in the
oscillation amplitude. It is also important to note that when
replacing neutrinos with antineutrinos the \dcp is replaced with
$-\dcp$. The effects of \dcp to \pr{\num}{\nue} are illustrated in
Fig.~\ref{fig:pmue}.
Considering these issues it can be seen that \dcp could be measured by
an experiment using the \ac*{JPARC} beamline if it could observe both
the first and second oscillation maximum, or run with both neutrino
and anti-neutrino beams. Both maximums can be observed by varying
$E/L$, by either using two different baselines $L_1$ and $L_2$ or by
varying the energy --- recalling that the neutrino energy increases
(and becomes less monochromatic) as the off-axis angle decreases. With
this in mind it is instructive to consider Fig.~\ref{fig:map}
showing a map with off-axis angle contours for the \ac*{JPARC}
neutrino beam on the surface of the earth.
\subsection{Okinoshima}
\begin{figure}[htbp]
  \vspace*{1\baselineskip}
  \includegraphics[width=0.49\columnwidth,height=!]{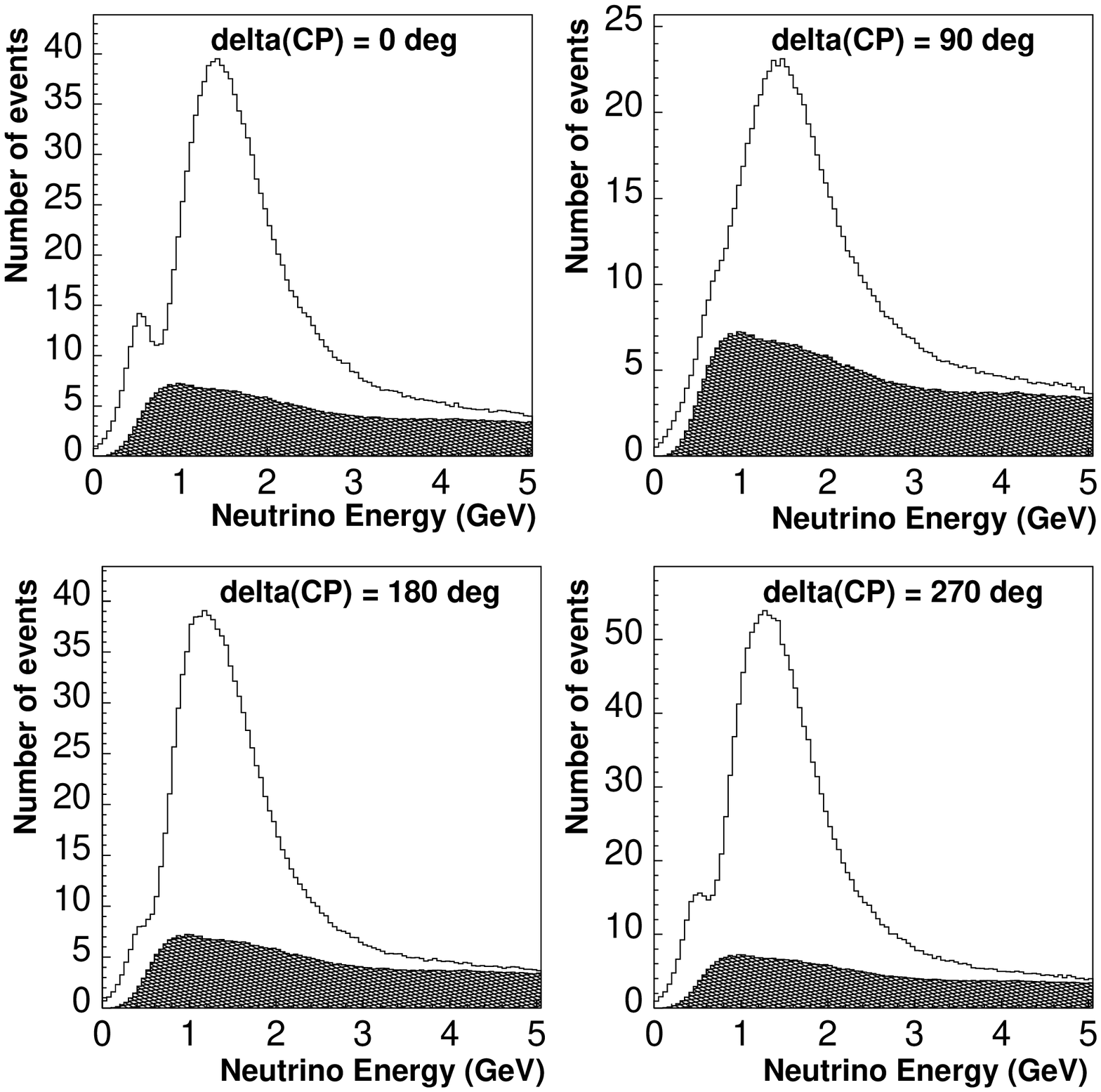}
  \includegraphics[width=0.49\columnwidth,height=!]{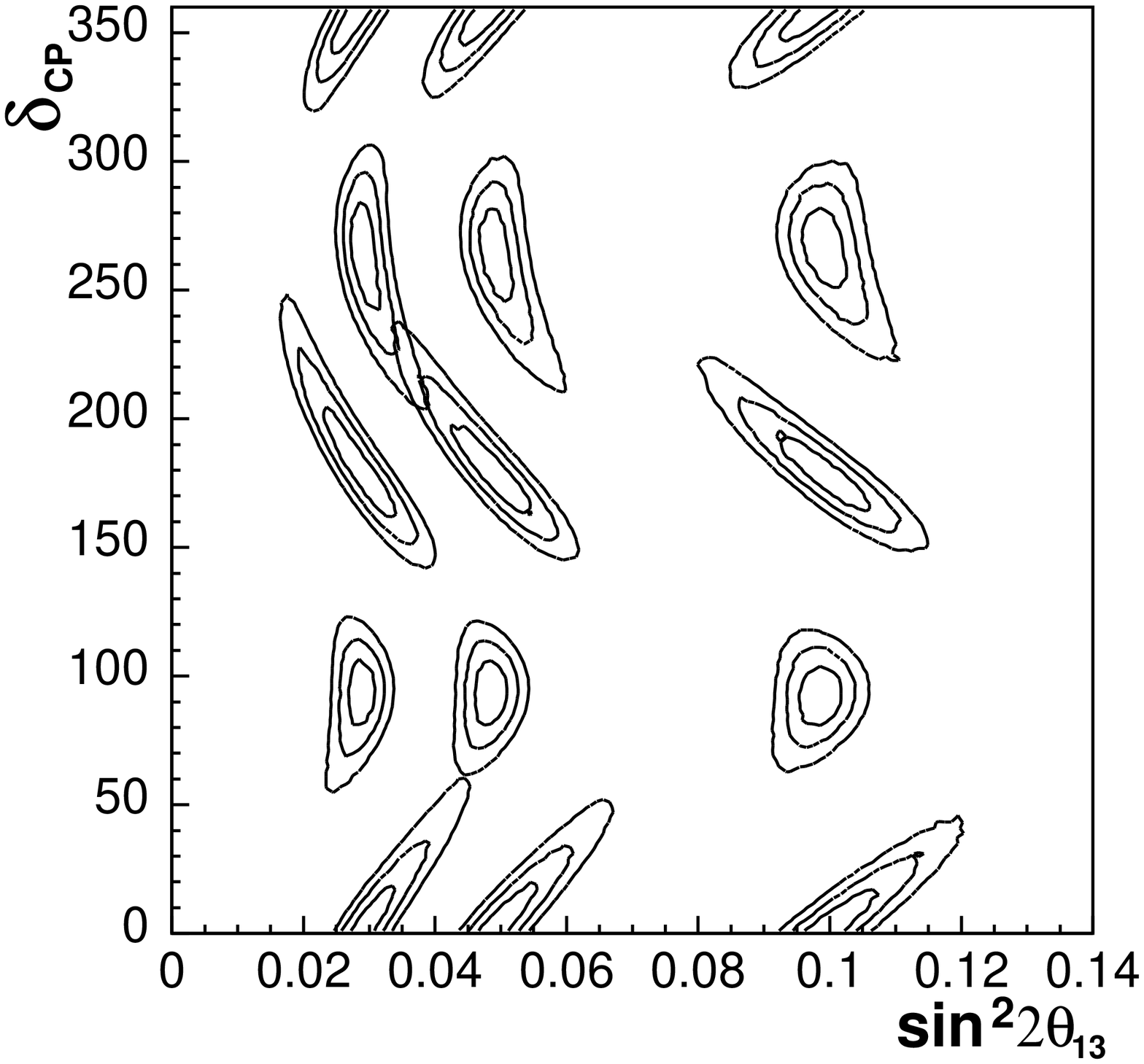}
  \caption{\label{fig:okinoshima}Very preliminary Okinoshima
    simulation results.  The left plots show the energy spectra for
    (clockwise from top left) $\dcp=0\degree$, $90\degree$,
    $270\degree$ and $180\degree$.  The right plot shows C.L.=67\%,
    95\% and 99.7\% allowed regions in the \sinstt{13}-\dcp plane.
    Twelve allowed regions are overlaid for the twelve combinations
    of $\sinstt{13}=0.1$, $0.05$, $0.02$ and
    $\dcp=0\degree$,$90\degree$, $270\degree$ and $180\degree$.\cite{Badertscher:2008bp}}
\end{figure}

Figure~\ref{fig:map} shows the island of Okinoshima
658~\kilo\metre\ from \ac*{JPARC} at a small off-axis angle of about
0.76\degree. The small off-axis angle provided by this location
results in significant neutrino beam energy spread allowing a detector
to observe not only the first but also the second oscillation maximum
assuming that it had sufficiently good energy resolution below
1~\gev. As such it is proposed to place a Liquid Argon
\ac*{TPC}\cite{Rubbia:2004tz} on
Okinoshima\cite{Badertscher:2008bp}. Such a detector is a precision
device with excellent energy and spacial resolution. The spatial
resolution that could be achieved would be able to significantly
suppress the dominant \pizero background in \nue reconstruction.
Simulations with a 100~\kilo\ton\ fiducial mass, $\sigma(E_\nu) =
100~\mev$, 5 years with a \num beam and an input value of
\sinstt{13}=0.03 yields the distributions and allowed regions in the
\sinstt{13}-\dcp plane shown in Fig.~\ref{fig:okinoshima}.
\subsection{Kamioka: ``Hyper-K''}
\begin{figure}[htbp]
  \includegraphics[width=1.0\columnwidth,height=!]{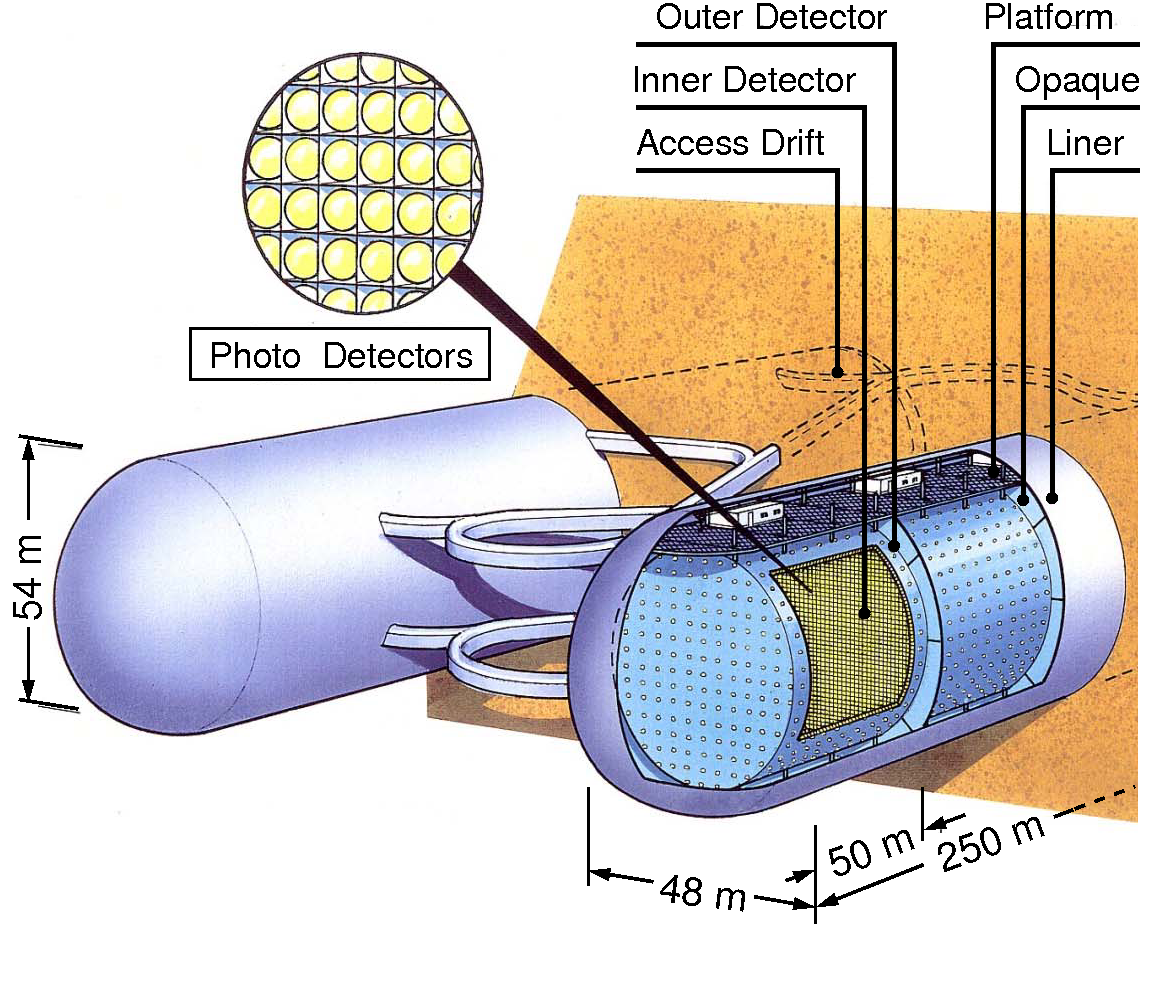}
  \caption{\label{fig:hk}The proposed Hyper-K water Cerenkov detectors.}
\end{figure}

\begin{figure}[htbp]
  \resizebox{1.0\columnwidth}{!}{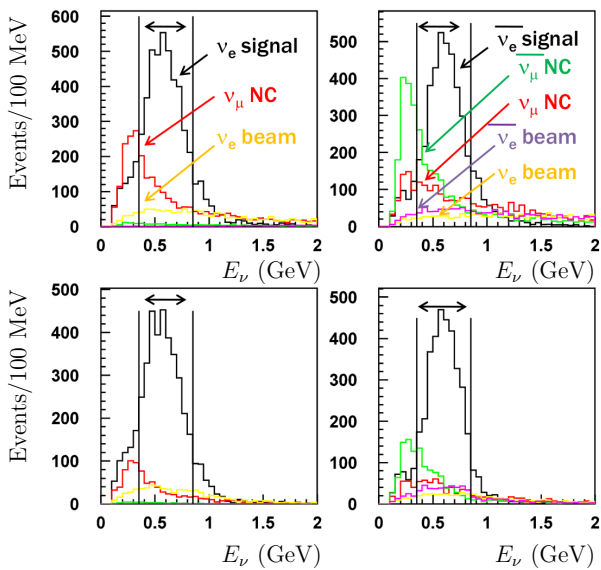}
  \caption{\label{fig:t2hk-dist}Hyper-K reconstructed energy
    distributions. The left plots are for \num beam and the right
    plots for \numb beam. The upper and lower plots show the
    distributions before and after final analysis cuts respectively.}
\end{figure}

\begin{figure}[htbp]
  \includegraphics[width=0.8\columnwidth,height=!]{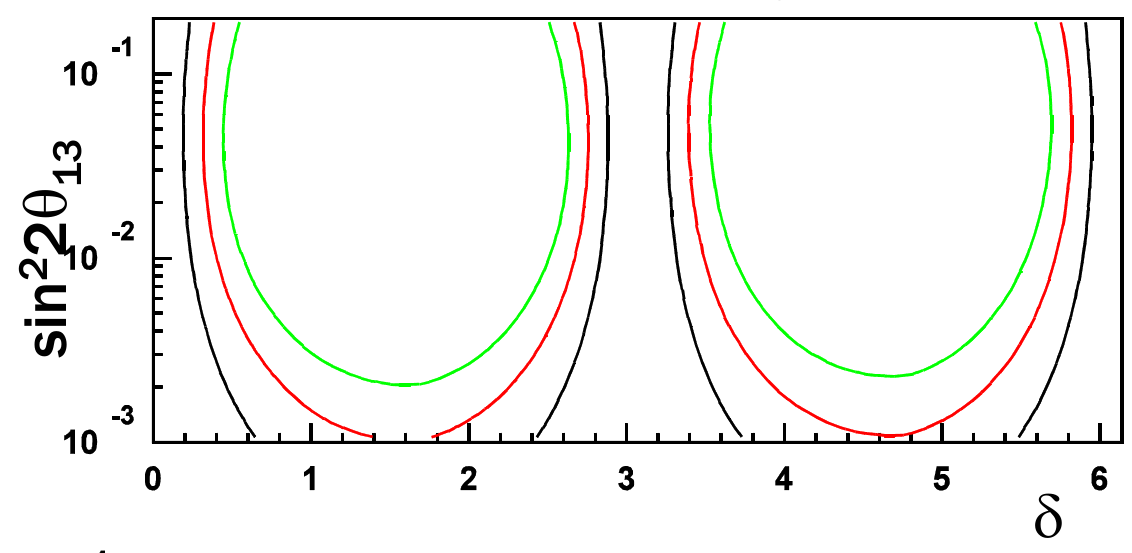}
  \caption{\label{fig:t2hk-exclu}Hyper-K \CP violation sensitivity
    plots. Contours indicate $1\sigma$, $2\sigma$ and $3\sigma$
    regions in the \sinstt{13}-\dcp plane.}
\end{figure}

The Hyper-K project is a plan to build a new 1000~\kilo\ton\ ($\simeq
500~\kilo\ton$ fiducial) volume Water Cerenkov detector at the Kamioka
site. The massive volume --- approximately 24 times larger that
\ac*{SK} --- would be realised using two 500~\kilo\ton\ chambers
illustrated in Fig.~\ref{fig:hk},
with a total of $0.2\times10^6$ \acp*{PMT}. Simulation studies\cite{NP08T2HK}
assuming 2.2 years of \num and 7.8 years of \numb running with
$\sinstt{13}=0.1$ yield the distributions in Fig.~\ref{fig:t2hk-dist}
and exclusion regions in Fig.~\ref{fig:t2hk-exclu}.
\subsection{Korea: ``T2KK''}
\begin{figure}[htbp]
  \includegraphics[width=0.49\columnwidth,height=!,trim= 0 15 55 80,clip]{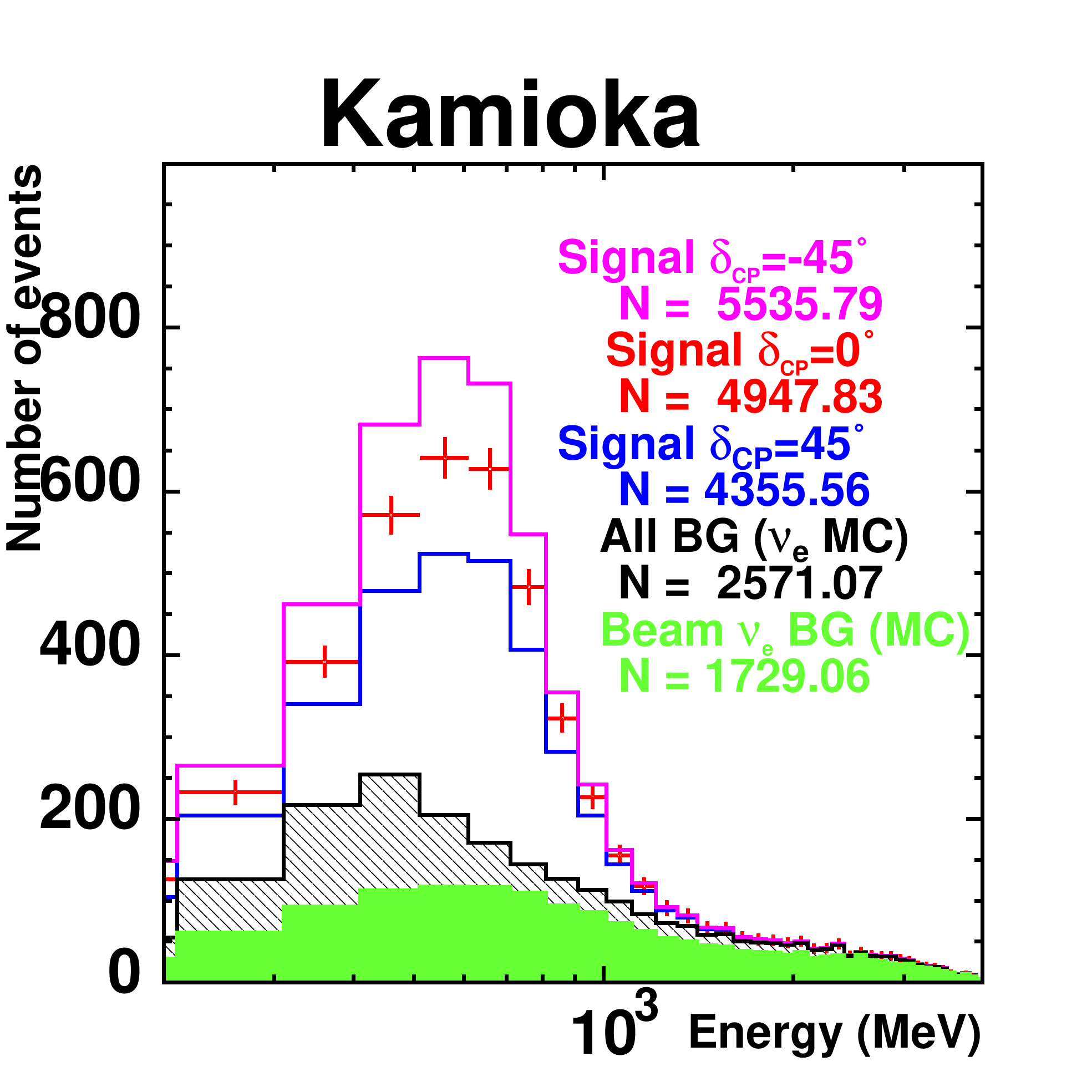}
  \includegraphics[width=0.49\columnwidth,height=!,trim= 0 15 55 80,clip]{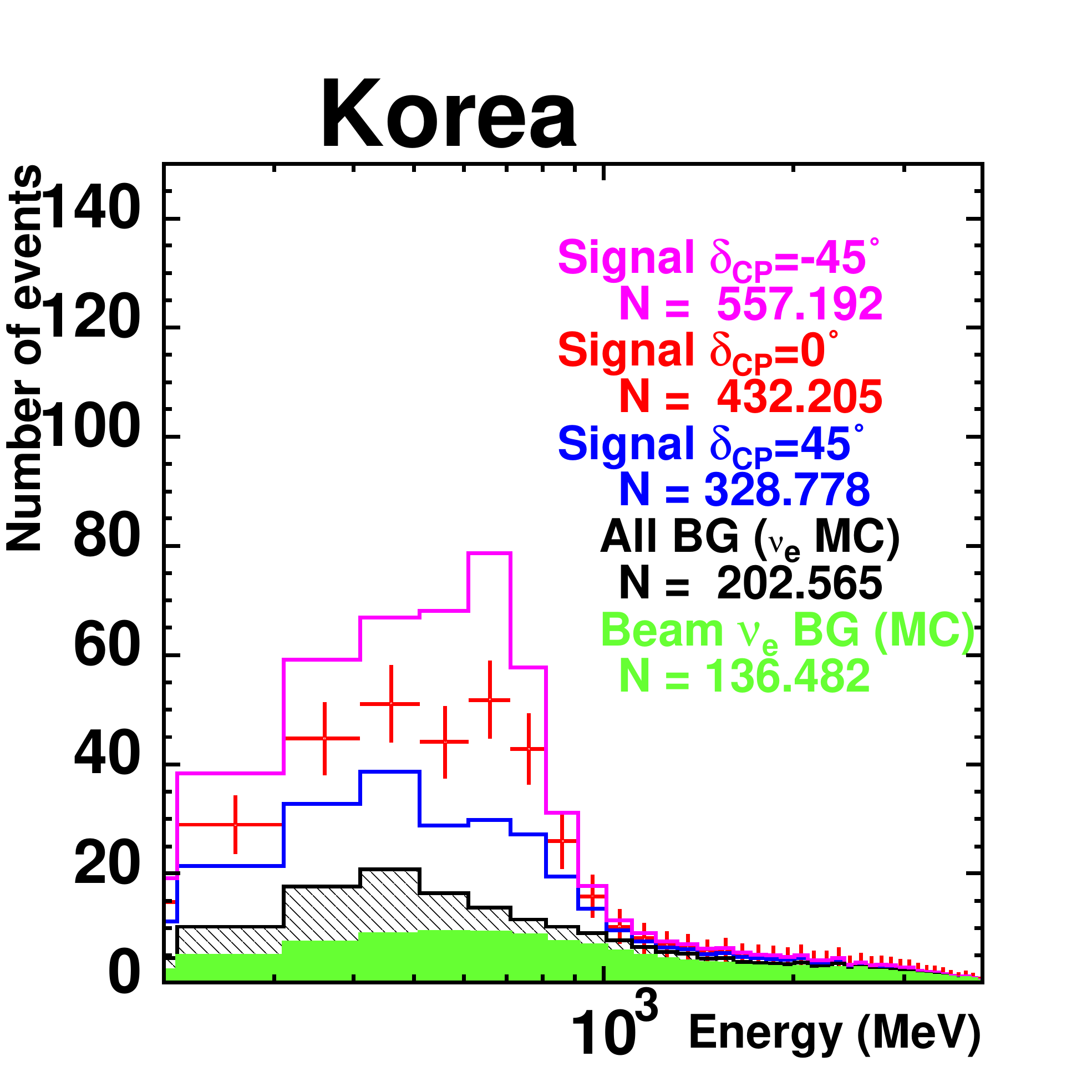}
  \caption{\label{fig:t2kk-dist}T2KK reconstructed neutrino energy
    distributions for Kamioka (left) and Korea (right).}
\end{figure}

\begin{figure}[htbp]
  \includegraphics[width=1.0\columnwidth,height=!,trim= 0 272 0 55,clip]%
      {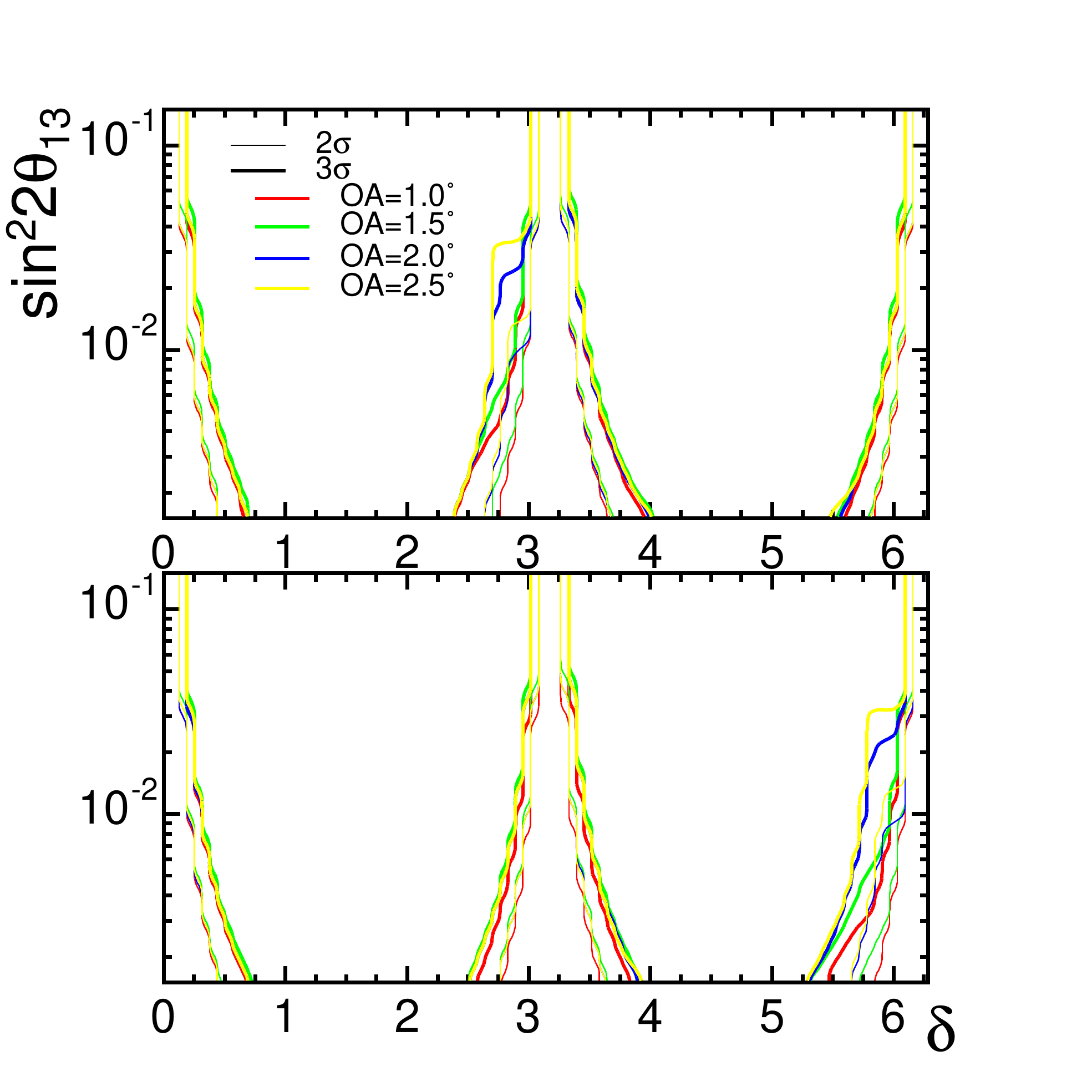}
  \caption{\label{fig:t2kk-alow}T2KK \CP violation sensitivity
    plots. Narrow and bold lines indicate $2\sigma$ and $3\sigma$
    regions. The different colours represent different off-axis angles
    and indicate no strong preference in choice of angle.}
\end{figure}

Inspection of Fig.~\ref{fig:map} shows that the \ac*{JPARC} neutrino
beam extends to South Korea with baselines from 1000~\kilo\metre\ to
1250~\kilo\metre\ and off-axis angles from 1.0\degree\ to
4.0\degree. Selecting a baseline and off-axis angle of
1000~\kilo\metre\ and 2.5\degree\ would place a detector at the second
neutrino oscillation maximum. Hence the idea is to place one
of the Hyper-K detectors at Kamioka and one in Korea. Preliminary
simulations\cite{NP08T2KK} with five years \num beam and five years \numb beam with an
input value of $\sinstt{13}=0.04$ yields the distributions in
Fig.~\ref{fig:t2kk-dist} and allowed regions in the \sinstt{13}-\dcp
plane shown in Fig.~\ref{fig:t2kk-alow}
\section{Summary}
The \ac*{T2K} experiment will utilise a new accelerator facility,
\ac*{JPARC} delivering statistics two orders of magnitude above those
of \ac*{K2K}. An off-axis beam configuration will be utilised
providing a narrow neutrino energy spectrum tuned to the oscillation
maximum. The beam will be monitored by an extensive suite of detectors
at the production site and will utilise the well established \ac*{SK}
as the far detector. \ac*{T2K} will search for \nue appearance with an
order magnitude improvement on $\sin^22\tmix{13}$ sensitivity over
current limits and will perform precision measurements of \num
disappearance with estimated uncertainties of
$\delta(\sin^22\tmix{23}) \simeq 0.01$ and $\delta(\dms{23}) <
10^{-4}~\electron\volt\squared$.  Construction of the \ac*{T2K}
facility at \ac*{JPARC} is well underway with neutrino beam
commissioning to begin in April 2009. If a sizable \decay{\num}{\nue}
signal is observed it is planned to upgrade the beam power and pursue
one of the promising detector options under study to pursue \CP
physics studies in the neutrino sector.

\bibliographystyle{unsrt}
\bibliography{neutrino}

\begin{acronym}
\acro{BPM}{Beam Position Monitor}
\acro{LPM}{Loop Pickuo Monitor}
\acro{ESM}{Electrostatic Monitor}
\acro{SSEM}{Segmented Seconday Emission Monitor}
\acro{CT}{Current Transformer}

\acro{NC}{Normal Conducting}
\acro{SC}{Super Conducting}
\acro{Prep}{Preparation}
\acro{FF}{Final Focus}
\acro{MS}{Monitor Stack}

\acro{KTK}[K2K]{KEK to Kamioka}
\acro{K2K}[K2K]{KEK to Kamioka}
\acro{T2K}{Tokai to Kamioka}
\acro{SK}{Super-Kamiokande}
\acro{ND}[ND280]{Near Detector at 280 metres}
\acro{TPC}{Time Projection Chamber}
\acro{KEK}{Japan High Energy Accelerator Laboritory}
\acro{JPARC}[J-PARC]{Japan Proton Accelerator Research Complex}

\acro{MNS}{Maki Nakagawa Sakata}
\acro{GUT}{Grand Unified Theory}
\acro{SM}{Standard Model of Particle Physics}
\acro{SSM}{Standard Solar Model}

\acro{CCQE}[CC-QE]{Charged Current Quasi-Elastic}
\acro{CCNQE}[CC-nQE]{Charged Current NonQuasi-Elastic}
\acrodef{PMT}{photomultiplier tube}
\acrodef{MR}{Main Ring}
\acrodef{RCS}{Rapid Cycling Syncrotron}
\end{acronym}          
\end{document}